# Solitary waves in Parity-time (PT) symmetric Bragg-grating structure and the existence of Optical Rogue Waves

Samit Kumar Gupta and Amarendra K. Sarma*
Department of Physics, Indian Institute of Technology Guwahati, Guwahati-781 039, Assam, India
*Electronic address: aksarma@iitg.ernet.in

In this work, we have studied the traveling wave solution in a nonlinear Bragg grating structure in which the core of the optical fiber is having Parity-time (PT) symmetric refractive index distribution. We have found bright solitary wave solution below the PT-threshold for forward wave and dark solitary wave solution above the PT-threshold for backward wave. The effects of increasing the traveling wave speed on the spatio-temporal evolutions of the analytical solutions have been shown and the emergence of the Optical Rogue Waves (ORWs) has been explored based on the system parameters.

## I. INTRODUCTION

Since the inception of the concept of non-Hermitian quantum mechanics built on the idea of parity-time symmetry [1], rigorous and widespread theoretical investigations have been carried out in a diverse areas of subjects pertaining to open quantum systems, quantum field theory and so on [2-6]. In recent years, the idea of parity-time symmetry has been experimentally observed in coupled-waveguide systems [7-8], synthetic photonic lattices [9] and complex optical potentials [10]. All these experimental investigations have greatly influenced the extension of the idea of parity-time symmetry in optics keeping in mind the huge potential applications that this novel concept can offer. In their path-breaking work, Bender and Boettcher have shown that for the eigen-values of the non-Hermitian Hamiltonian $\hat{H} = \hat{p}^2/2m + V(\hat{x})$, where m is the mass, $\hat{x}$ is the position operator, $\hat{p}$ is the momentum operator, V is the complex potential, to be entirely real, the Hamiltonian should share common set of eigen-vectors with the PT-operator such that [H, PT] = 0. P is the parity inversion operator defined as: $\hat{p} \to -\hat{p}, \hat{x} \to -\hat{x}$, and T is the time-reversal operator defined as: $\hat{p} \to -\hat{p}, \hat{x} \to \hat{x}, i \to -i$. It comes out that the Hamiltonian will be PT-symmetric only when $V^*(-x) = V(x)$ or equivalently $\text{Re}[V(-x)] = \text{Re}[V(x)]$ and $\text{Im}[V(-x)] = -\text{Im}[V(x)]$. Similarly, an optical system having its complex refractive index distribution as $n(x) = n_R(x) + i n_I(x)$ is said to be PT-symmetric when $n(x) = n^*(-x)$ or $n_R(x) = n_R(-x)$ and $n_I(-x) = -n_I(x)$, meaning that the refractive index distribution must be an even function of position whereas the gain/loss distribution must be odd. The inclusion of the PT-symmetry into a physical system can invoke new behaviors in its dynamical characteristics which otherwise would have been impossible for the case of passive optical systems. It is essentially this point which is drawing lots of research interests in recent times in terms of PT-symmetry. The notion of PT-symmetry has been explored by various research groups with reference to nonlinear optics. Some of them are: beam dynamics in PT-symmetric optical lattices [11], Anderson localization of light in PT-symmetric optical lattices [12], optical coupler with birefringent arms [13] and so on.

Again, periodic structures like index gratings in bulk [14] and in optical fibers [15-16], find important applications in obtaining frequency selectivity, high reflectivity and dispersion of significant characteristic properties [17]. In their work [18], M. A. Miri et al. starting from a modified massive Thirring model [19], have established the existence of slow Bragg solitons in the complex nonlinear parity-time (PT) symmetric periodic structures and the dependence of the grating band structures on the PT-symmetric part of the periodic optical refractive index. Yet in another work, N. A. Kudryashov et al. [20], in a system of double tunnel-coupled waveguides, have investigated the analytical traveling wave solutions including two periodic and solitary waves and a new nonlinear wave called compacton. Besides, solitary waves have been studied extensively in the oppositely directed couplers [21-23], in systems with different kinds of nonlinearities modeled by nonlinear Schrodinger equations (NLSEs) [24-32]. Recently, with detailed studies of "rogue waves" or "freak waves" in hydro-dynamical systems [33-35], the investigation of rogue waves has been ramified into nonlinear fiber optics [36] and Bose-Einstein Condensates (BECs) [37]. Optical Rogue Waves have been studied in a microstructured optical fiber system, near the threshold position of soliton-fission supercontinuum generation has been observed [38]. Even more recently, ORWs have been investigated in singly resonant optical parametric oscillators [39], optically injected lasers [40] and in mode-locked fiber lasers [41]. Motivated by these works, we consider a Bragg grating structure with core having refractive index distribution as indicated in the section II. For this system we study the traveling wave solutions of the system depending on the system parameters and explore the nature of the solutions.

The article is organized as follows: Section II briefly describes the theoretical model. Section III discusses the analytical traveling wave solutions. The nature of the solutions below and above the PT-symmetric threshold is illustrated. Finally, the existence of optical rogue waves is discussed in Section IV followed by conclusions in Section V.

## II. THEORETICAL MODEL

We are considering a nonlinear Bragg grating structure in which the optical fiber-core is associated with the core refractive index distribution as given by: $n = n_0 + n_{1R}\cos(2\pi z/\Lambda) + i n_{1I}\sin(2\pi z/\Lambda) + n_2 |E|^2 E$. Here, $n_0$ is the refractive index of the background material; $n_{1R}$ and $n_{1I}$ are the real and imaginary parts of the perturbed refractive index distribution; $n_2$ is the nonlinear Kerr parameter, while $\Lambda$ is the grating period. The equations, in the slowly-varying envelope approximation describing the evolutions of the forward and backward waves are given below:

$$i\left(\frac{\partial E_f}{\partial z} + \frac{1}{v_g}\frac{\partial E_f}{\partial t}\right) + \delta_0 E_f + (k_0 + g_0)E_b + \gamma_0(|E_f|^2 + 2|E_b|^2)E_f = 0$$

$$i\left(\frac{\partial E_b}{\partial z} - \frac{1}{v_g}\frac{\partial E_b}{\partial t}\right) - \delta_0 E_b - (k_0 - g_0)E_f - \gamma_0(|E_b|^2 + 2|E_f|^2)E_b = 0$$

(1)

Here $E_f(z,t)$ and $E_b(z,t)$ are the slowly-varying amplitudes of the forward and the backward waves respectively. $v_g = c/n_0$ is the velocity in the background material, $k_0 = \pi n_{1R}/\lambda_0$ is the coupling co-efficient arising from the real Bragg grating and $g_0 = \pi n_{1I}/\lambda_0$ is the anti-symmetric coupling co-efficient arising from the complex PT-potential term. $\delta = (\omega_0 - \omega_B)/v_g$ is the measure of the detuning from the Bragg frequency, $\omega_B$. $\gamma_0$ is the so called self-phase modulation parameter. In order to simplify our analysis, we adopt the following dimensionless units:

$$\xi = z/z_0, \tilde{E}_f = E_f/\sqrt{P_0}, \tilde{E}_b = E_b/\sqrt{P_0}, \tau = t/T_0, \delta = \delta_0 z_0, k = k_0 z_0, k = k_0 z_0, \gamma = \gamma_0 P_0 z_0.$$

Now, Eq.(1) can be rewritten as follows:

$$i\left(\frac{\partial \tilde{E}_f}{\partial \xi} + \frac{\partial \tilde{E}_f}{\partial \tau}\right) + \delta \tilde{E}_f + (k+g)\tilde{E}_b + \gamma(|\tilde{E}_f|^2 + 2|\tilde{E}_b|^2)\tilde{E}_f = 0$$

$$i\left(\frac{\partial \tilde{E}_b}{\partial \xi} - \frac{\partial \tilde{E}_b}{\partial \tau}\right) - \delta \tilde{E}_b - (k-g)\tilde{E}_f - \gamma(|\tilde{E}_b|^2 + 2|\tilde{E}_f|^2)\tilde{E}_b = 0$$

(2)

It is noteworthy that in deriving the Eq. (2), the coupled-mode theory is employed which has been used with considerable success in diverse contexts [38]. In this method, the forward and the backward waves are considered independently and the Bragg grating provides a coupling bridge between them.

### III. ANALYTICAL TRAVELLING WAVE SOLUTIONS

To start with, we make the transformation: $\tilde{E}_{f,b}(\xi,\tau) = a_{f,b}(\xi,\tau) e^{i\phi_{f,b}(\xi,\tau)}$, $\phi(\xi,\tau) = \phi_f - \phi_b$, in Eqs. (2). This results in the following set of equations for the real amplitudes $a_{f,b}(\xi,\tau)$ and real phases $\phi_{f,b}$:

$$\left(\frac{\partial}{\partial \xi} + \frac{\partial}{\partial \tau}\right)a_f = \sin(\phi)(k+g)a_b$$

$$\left(\frac{\partial}{\partial \xi} - \frac{\partial}{\partial \tau}\right)a_b = \sin(\phi)(k-g)a_f$$

$$a_f\left(\frac{\partial}{\partial \xi} + \frac{\partial}{\partial \tau}\right)\phi_f = \delta a_f + \cos(\phi)(k+g)a_b + \gamma(a_f^3 + 2a_f a_b^2)$$

$$a_b\left(\frac{\partial}{\partial \xi} - \frac{\partial}{\partial \tau}\right)\phi_b = -\delta a_b - \cos(\phi)(k-g)a_b - \gamma(a_b^3 + 2a_b a_f^2)$$

(3)

Now, we introduce the following traveling wave variables in Eq. (3):

$$z = (\xi - c_0 \tau)/\sqrt{1-c_0^2}, \quad a_f = \sqrt{1+c_0}\, u, \quad a_b = \sqrt{1-c_0}\, v \tag{4}$$

In order to keep the rest of the analysis clean, we introduce the following parameters:

$$\alpha = k+g, \beta = k-g, \mu = \sqrt{1+c_0 / 1-c_0} \qquad (5)$$

Using Eq. (4) in Eq. (3) and after doing some straightforward algebra, we obtain:
$u^2 = v^2(\alpha/\beta) + c_1$, where, $c_1$ is an integration constant. (6)

For $c_1 \neq 0$, from Eqs. (3), (4) and (5) we obtain:

$$u_z^2 = (u^2 + c_1)\sin^2\phi / \beta^2, \quad \phi_z = \alpha\varepsilon\left[u_{zz}(u^2 - c_1) - uu_z^2\right] / \chi(u^2 - c_1)\sqrt{u^2 - c_1 - u_z^2} \qquad (7)$$

where, $\varepsilon = \pm 1, \chi = \sqrt{\beta/\alpha}$ and $\phi = \sin^{-1}\left[\alpha\varepsilon u_z / \chi\sqrt{u^2 - c_1}\right]$ (8)

Substituting Eq. (8) into Eq. (7) and using the new variables, defined as, $E = \left[u^2 - c_1 - u_z^2\alpha/\chi\right]/u^6$ and $r = \sqrt{E}$, we obtain:

$$\frac{dr}{du} + \frac{4}{u}r = \varepsilon\alpha\frac{b_1 u^2 + b_2}{u^3} \qquad (9)$$

where, $b_1 = \gamma[\chi^2(\mu^{-1} + 2\gamma\sqrt{1-c_0^2}) + \mu(1+c_0)][\chi^2(\mu^{-1} + 2\gamma\sqrt{1-c_0^2}) + \mu(1+c_0)]$ and $b_2 = \gamma' + \gamma[-c_1\chi^2(\mu(1+c_0) + 2\sqrt{1-c_0^2})]$ with $\gamma' = \delta(\mu + \mu^{-1})$.

Solving Eq. (9) and then putting $w = 1/u^2$, after a bit of tedious algebra, under the approximation of weak nonlinearity, we get:

$$w_z^2 = [-4c_2^2 w^4 - 4c_1 w^3 + 4w^2 - 2c_2\varepsilon\alpha(b_1/2 + 2b_2 w/3)](\chi/\alpha)^2 \qquad (10)$$

Now, we write the right hand side of Eq. (10) as:
$f(w) = -2(w - w_1)(w - w_2)(w - w_3)(w - w_4)$,

Then, proceeding in the similar way as in Refs. [20, 43], we finally obtain the travelling wave solutions for the forward and backward waves:

$$a_f^2 = (1+c_0)\frac{\left[\rho + \tanh^2 Z\right]}{\left[w_4 + w_3 \tanh^2 Z\right]}$$

$$a_b^2 = (1-c_0)\left(\frac{k-g}{k+g}\right)\left(-c_1 + \frac{a_f^2}{1+c_0}\right)\frac{\left[\rho + \tanh^2 Z\right]}{\left[w_4 + w_3 \tanh^2 Z\right]} \qquad (11)$$

with $Z = \sqrt{3}\left(\sqrt{l}z + c\right)$. Here, '$f$' stands for forward wave and '$b$' stands for backward wave and '$c$' is the integration constant. '$l$' is the parametric constant defined as: $l = w_4^2(1 - X_1)(1 - X_2)/2$, where $X_m = w_m/w_4$, m=1,2,3,4; and $\rho = -w_3/w_4, c_1 = \rho/w_4$. It is clear that there are an infinite number of traveling wave solutions depending on the values of the parameters. One may obtain either cnoidal wave or solitary wave solutions subject to the parameter values. For the rest of our analysis in this work we take $c = 0.0$.

It is worthwhile to mention that at k=g, solitary wave solutions as in Eq. (11) do not exist for the chosen set of parameters. This could be due to the fact that in this case the

Bragg grating becomes ineffective since the speed reduction factor tends to unity [42]. Also, equivalently, as evident from the Eq. (6), at k=g, the solutions simply blows up.

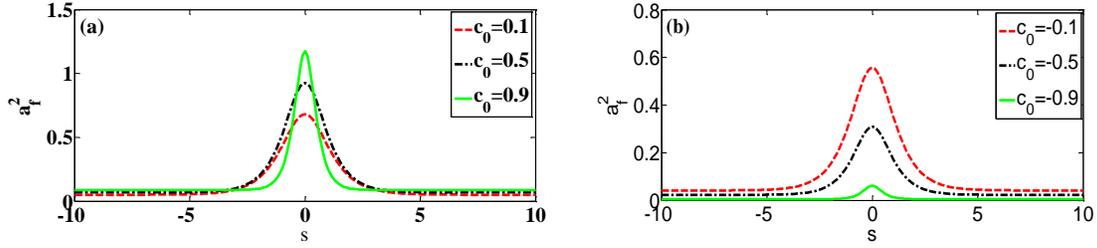

FIG. 1. (Color online) Variation of the square of the amplitudes vs. the *s* parameter for different values of $c_0$ parameter. k=1.0, g=0.9.

The variation of $a_f^2$ with $s = \xi - c_0 \tau$, for different values of $c_0$ parameters is shown in Fig.1. We notice that for the right-going waves ($0 < c_0 < 1$), the amplitudes increase when the traveling wave speed $c_0$ increases. For the left-going waves, the amplitudes decrease when the absolute values of the traveling wave speed is increased. These results are in conformity with the ones found in Refs. [20,22].

## IV. BRIGHT AND DARK SOLITARY WAVES AND THE EMERGENCE OF OPTICAL ROGUE WAVES FOR PARTICULAR SOLUTIONS

We mentioned in Sec. III that there are many possible solutions to Eq. (11) subject to various parameter values. One interesting solution may be of the so called Optical Rogue Waves (ORW). In fact, Eq. (11) has close similarity with the solutions reported recently in the context of nonlinear Fiber optics [44]. In this section, we discuss the spatio-temporal evolutions of the forward and the backward waves as found in Eq. (11).

Fig. 2 and 3 depicts the spatio-temporal evolution of the intensity for forward and backward waves respectively, below the PT-threshold.

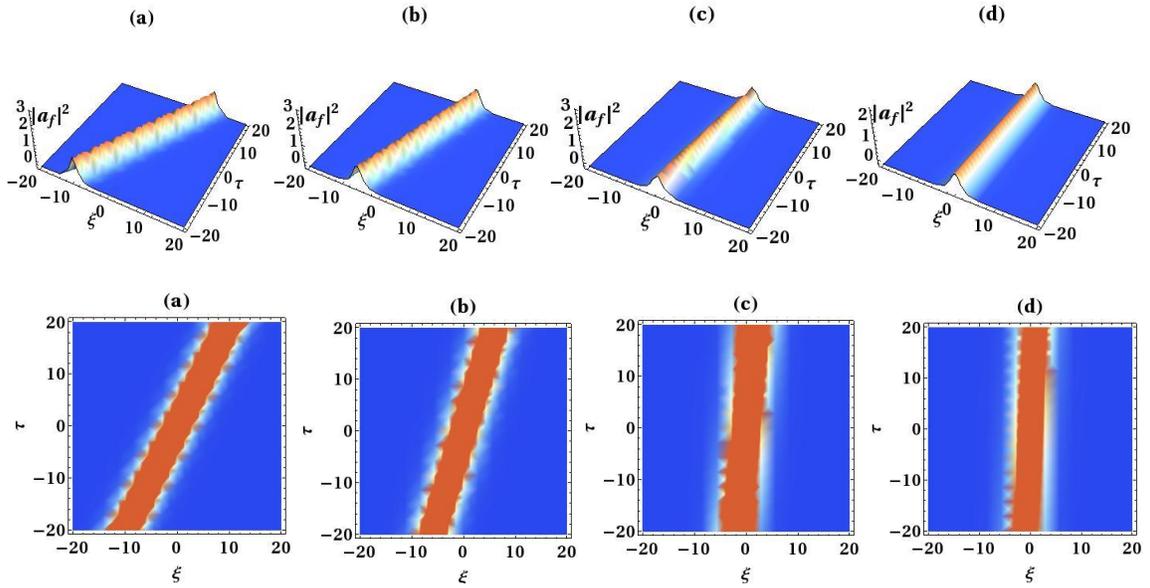

FIG. 2.(Color online) Spatio-temporal evolution of the intensity of the forward waves in the upper panel. In the lower panel density-plots corresponding to the upper panel is shown. Different values of $c_0$ parameter are: (a) $c_0=0.5$, (b) $c_0=0.3$, (c) $c_0=0.09$ and (d) $c_0=0.046$. Other parameters are: k=1.0 and g=0.55.

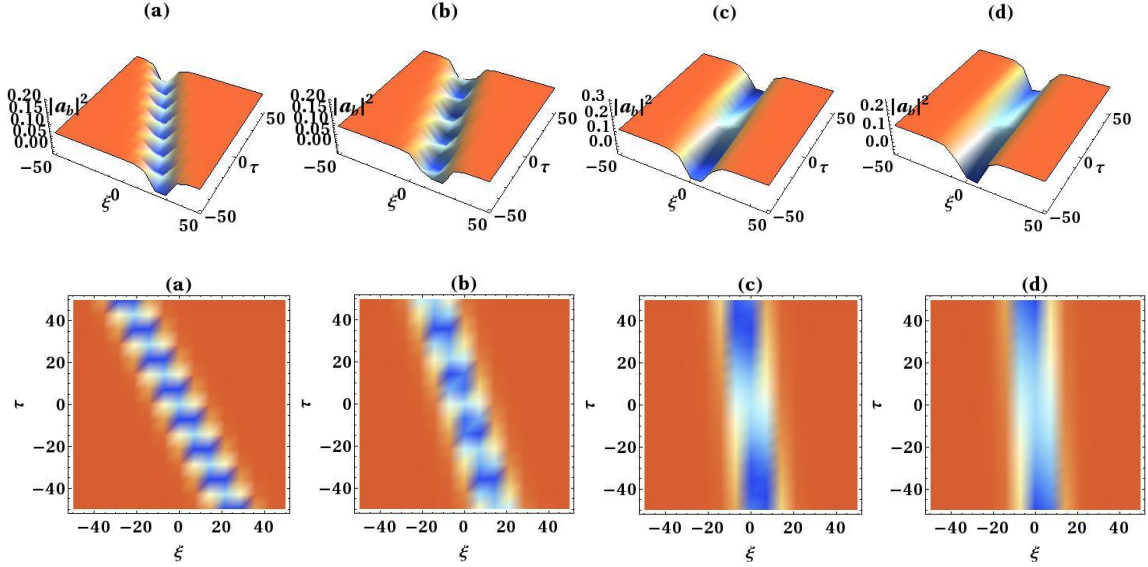

FIG. 3.(Color online) Spatio-temporal evolution of the intensity of the backward waves below the PT-threshold (upper panel). The corresponding density-plots in the lower panel. Different values of $c_0$ parameter are: (a) $c_0=-0.5$, (b) $c_0=-0.3$, (c) $c_0=-0.09$ and (d) $c_0=-0.046$. Other parameters are: k=1.0, g=0.90.

The simulations are carried out for two set of values of the traveling wave speed $c_0$: one with positive and the other one with negative values. The positive values of $c_0$ refers to forward wave while the negative ones refers to backward waves. It is quite interesting to find that, below the PT-threshold the system gives rise to bright solitary wave for forward wave while dark solitary wave solution for backward wave for a chosen set of parameters. This could be clearly observed from Figs. 2 and 3. It should be noted that stable solitary wave propagation occurs for small value of $c_0$ below the PT-threshold. Also, the solitary wave is shifted towards the positive $\xi$ axis for forward wave while the opposite occurs for backward wave with increase in the value of $c_0$. The bright as well as dark solitary waves get deformed with increase in the value of the traveling wave speed.

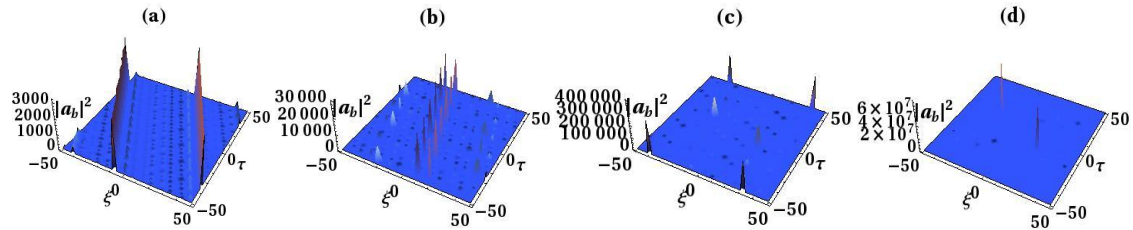

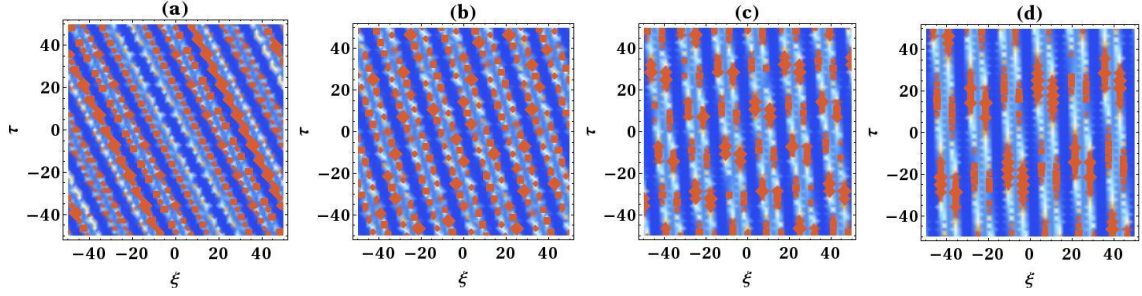

FIG. 4.(Color online) Spatio-temporal evolution of the intensity for backward waves above the PT-threshold (upper panel). In the lower panel density-plots corresponding to the upper panel is shown. Different values of $c_0$ parameter are: (a) $c_0$=-0.5, (b) $c_0$=-0.3, (c) $c_0$=-0.09 and (d) $c_0$=-0.046. Other parameters are: k=1.0, g=1.2.

The spatio-temporal evolution of the intensity for backward waves above the PT-threshold, i.e. the case with k<g, is depicted in Fig. 4. Solitary wave with high intensity could be observed from Fig. 4(a). If the modulus of the $c_0$ parameter is decreased significantly then it results in the generation of extremely high intensity backward wave. In fact, Fig. 4(c) and (d) predicts the existence of localized, single solitary waves of large amplitude in the background medium. These occurrences might be explained as follows. As the modulus of the traveling wave speed $c_0$ is decreased by a little amount, it acts as a perturbation. The perturbation gets amplified via nonlinear processes such as modulation instability, which in turn decreases the soliton-fission threshold. Now, as the soliton-fission sets in, it results in the breaking up of the smooth wave and the formation of well-localized group of solitary waves of large amplitudes resembling closely to the so called optical Rogue waves. Physically speaking, with decrease in the modulus of the traveling wave speed, the energy of each relatively small-amplitude localized wave starts transferring energy to another one amongst the group of those waves, thus increasing its amplitude rapidly. Interestingly, solutions cease to exist for the forward wave whereas the optical rogue wave solutions are obtained for backward wave above the PT-threshold region. It may be reasonable to say that the absence of the forward wave and the presence of the backward wave is complementary to each other.

In passing, it may be appropriate to note that in this work we have assumed the nonlinearity to be weak. Its effects are three-fold: firstly, the approximation of weak nonlinearity is ensuring the simple analytical solutions found in Eq.(11), secondly, since the nonlinearity is weak, it is practically not changing the position of the PT-threshold and third, the condition of weak nonlinearity gives rise to modulation instability accounting for the emergence of ORWs. For regular waves, Benjamin-Feir instability [45] is a weakly-nonlinear effect finding equivalent analogues in nonlinear optics. Besides, the weakly-nonlinear theory of the occurrence of rogue waves has been widely suggested as possible explaining tool [46, 46], although controversial. As regards some possible applications of the present work, many may be suggested, but one potential application may be to use this PT- symmetric Bragg grating structure in generating high power optical pulses [48].

## V. CONCLUSIONS

In this work, we have found the analytical traveling solitary wave solutions for the forward and the backward waves in a nonlinear PT-symmetric Bragg grating structure. We observe that there is an infinite number of traveling wave solutions, either solitary or cnoidal-type, depending on the parameter values of the system. The effects of the coupling constant, gain/loss parameter and the traveling wave speed on the evolution dynamics of the solutions have been discussed. We have found bright solitary wave solutions below the PT-threshold for forward wave, and dark solitary wave solutions above the PT-threshold for backward wave. Depending on some suitable choice of the parameter values of the system, the existence of optical rogue waves has been elucidated. It comes out that the evolutions of the forward and the backward waves crucially depend upon the different system parameters such as traveling wave speed and loss/gain parameter. It is worth mentioning here that the theoretical set-up under consideration can yield, apart from ORWs, various kinds of nonlinear traveling wave solutions based on the parameter values of the system. So, in that regard, this work could a step forward to the future investigations in the similar systems.

## ACKNOWLEDGMENTS

S.K.G. would like to thank, MHRD, Government of India, for a research fellowship.
-----------------------------------------------------------------------------------------------------------------